\documentclass{article}

\usepackage{microtype}
\usepackage{graphicx}
\usepackage{subfigure}
\usepackage{booktabs} 

\usepackage{xurl}
\usepackage{hyperref}

\usepackage[accepted]{icml2024}

\usepackage{amsmath}
\usepackage{amssymb}
\usepackage{mathtools}
\usepackage{amsthm}

\usepackage[capitalize,noabbrev]{cleveref}

\theoremstyle{plain}

\theoremstyle{definition}

\theoremstyle{remark}

\usepackage[textsize=tiny]{todonotes}

\usepackage{color}

\newcommand{\gfm}{\texttt{\textbf{CaPS}}}
\newcommand{\pSEL}{\ensuremath{\pi_{\mathsf{SELECT}}}}
\newcommand{\pMSR}{\ensuremath{\pi_{\mathsf{MEASURE}}}}
\newcommand{\pEXP}{\ensuremath{\pi_{\mathsf{EXP}}}}

\newcommand{\pCOMP}{\ensuremath{\pi_{\mathsf{COMP}}}}
\newcommand{\pLT}{\ensuremath{\pi_{\mathsf{LT}}}}
\newcommand{\pRDM}{\ensuremath{\pi_{\mathsf{GR-RANDOM}}}}
\newcommand{\pRDMB}{\ensuremath{\pi_{\mathsf{GR-RNDM-BIT}}}}

\newcommand{\pMUL}{\ensuremath{\pi_{\mathsf{MUL}}}}

\newcommand{\pLN}{\ensuremath{\pi_{\mathsf{LN}}}}
\newcommand{\pEQ}{\ensuremath{\pi_{\mathsf{EQ}}}}

\newcommand{\pGT}{\ensuremath{\pi_{\mathsf{GT}}}}

\newcommand{\pERR}{\ensuremath{\pi_{\mathsf{ERR}}}}
\newcommand{\pNORM}{\ensuremath{\pi_{\mathsf{NORM}}}}
\newcommand{\pPROB}{\ensuremath{\pi_{\mathsf{PROB}}}}
\newcommand{\pRC}{\ensuremath{\pi_{\mathsf{RC}}}}
\newcommand{\pMAX}{\ensuremath{\pi_{\mathsf{MAX}}}}
\DeclareMathOperator{\sgn}{sgn}

\newcommand{\pJOIN}{\ensuremath{\pi_{\mathsf{JOIN}}}}
\newcommand{\pLOG}{\ensuremath{\pi_{\mathsf{LN}}}}
\newcommand{\pSQRT}{\ensuremath{\pi_{\mathsf{SQRT}}}}
\newcommand{\pCOS}{\ensuremath{\pi_{\mathsf{COS}}}}
\newcommand{\pSIN}{\ensuremath{\pi_{\mathsf{SIN}}}}
\newcommand{\pGSS}{\ensuremath{\pi_{\mathsf{GSS}}}}
\newcommand{\pSOFTMAX}{\ensuremath{\pi_{\mathsf{SOFTMAX}}}}
\newcommand{\pSUM}{\ensuremath{\pi_{\mathsf{SUM}}}}
\newcommand{\pNORML}{\ensuremath{\pi_{\mathsf{L1-NORM}}}}
\newcommand{\pGTE}{\ensuremath{\pi_{\mathsf{GTE}}}}
\usepackage{xspace}

\newenvironment{myprotocol}[1][htb]{%
    \floatname{algorithm}{Protocol}
   \begin{algorithm}[#1]%
   \footnotesize{}
  }{\end{algorithm}}

\usepackage{multirow}
\usepackage{booktabs}
\usepackage{soul}
\usepackage{enumitem}

\newlength\myindent
\icmltitlerunning{}

\begin{document}

\twocolumn[
\icmltitle{CaPS: Collaborative and Private \\ Synthetic Data Generation from Distributed Sources}





\begin{icmlauthorlist}
\icmlauthor{Sikha Pentyala}{uwt}
\icmlauthor{Mayana Pereira}{ub}
\icmlauthor{Martine De Cock}{uwt,ghent}
\end{icmlauthorlist}

\icmlaffiliation{uwt}{School of Engineering and Technology, University of Washington Tacoma, USA}
\icmlaffiliation{ub}{Department of Electrical Engineering, Universidade de Brasilia, Brazil}
\icmlaffiliation{ghent}{Department of
Applied Mathematics, Computer Science and Statistics, Ghent University, Belgium}

\icmlcorrespondingauthor{Sikha Pentyala}{sikha@uw.edu}



\icmlkeywords{Secure Multiparty Computation, Synthetic Data, Privacy}

\vskip 0.3in
]



\printAffiliationsAndNotice{}  

\begin{abstract}
Data is the lifeblood of the modern world, forming a fundamental part of AI, decision-making, and research advances.
With  increase in interest in data, governments have taken important steps towards a regulated data world, drastically impacting data sharing and data usability and resulting in massive amounts of data confined within the walls of organizations. 
While synthetic data generation (SDG) is an appealing solution to break down these walls and enable data sharing, the main drawback of existing solutions is the assumption of a trusted aggregator for generative model training. 
Given that many data holders may not want to, or be legally allowed to, entrust a central entity with their raw data, we propose a framework for collaborative and private generation of synthetic tabular data from distributed data holders. 
Our solution is general, applicable to any marginal-based SDG, and provides input privacy by replacing the trusted aggregator with secure multi-party computation (MPC) protocols and output privacy via differential privacy (DP). 
We demonstrate the applicability and scalability of our approach for the state-of-the-art select-measure-generate SDG algorithms MWEM+PGM and AIM.
\end{abstract}

%
%
\section{Introduction}
\label{sec:intro1}

The success of data-driven applications in a variety of domains -- including but not limited to healthcare, finance, and automation -- can be attributed to the availability of data sharing for analysis. However, such sharing of data raises many privacy concerns. 
Laws and guidelines to protect user privacy in AI applications -- such as the GDPR\footnote{European General Data Protection Regulation\\ \url{https://gdpr-info.eu/}} in Europe and the AI Bill of Rights\footnote{\url{https://www.whitehouse.gov/ostp/ai-bill-of-rights/}} in the United States -- impose substantial barriers to data sharing across organisations, especially in highly regulated domains like healthcare and finance. 
\begin{figure}[ht!]
    \centering
    \includegraphics[width=8.5 cm]{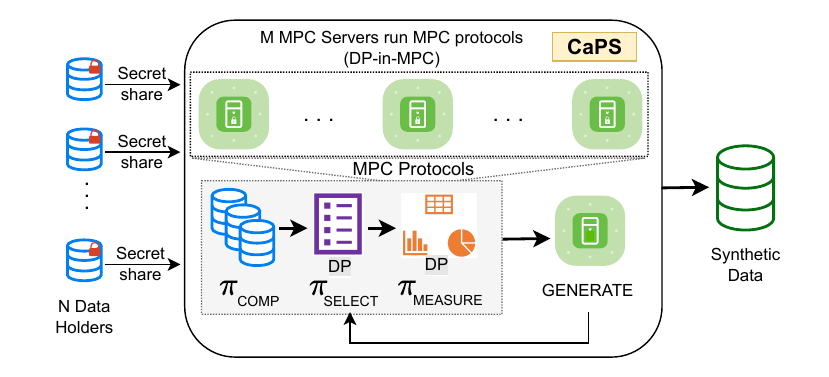}
    \caption{$\gfm$: A framework that leverages `DP-in-MPC' to collaboratively and privately generate tabular synthetic data using marginal-based SDG techniques with the `select-measure-generate' template. Servers run MPC protocols for `select' and `measure'. The `generate' step is performed over differentially private measurements. 
    }
    \label{fig:intro}
\end{figure}


\vspace{-1em}
Synthetic data, i.e.~artificial data generated using a purpose-built synthesizer trained on real data, offers an appealing solution to share microdata while mitigating privacy concerns. Synthetic data generation (SDG) with differential privacy (DP)
has received considerable attention from the research community (see e.g.~\cite{zhang2017privbayes, jordon2019pate, mckenna2021winning, tao2021benchmarking} and references therein). This generated data fits any data processing workflow designed for the original data,
while providing formal privacy guarantees.

Tabular data is ubiquitous in many domains ranging from healthcare, to humanitarian action, education, and socioeconomic studies 
Differentially private synthetic tabular data can be used for machine learning tasks and data analysis, as well as enable answering an arbitrary number of statistical queries. 
While many methods have been proposed in the literature to generate differentially private synthetic tabular data (DP-SD), such as GAN-based (e.g.~\cite{jordon2019pate,xu2019modeling}) and marginal-based (e.g.~\cite{mckenna2021winning,mckenna2022aim}) approaches, benchmarking studies have shown that marginal-based synthetic data generators often generate high quality tabular data with strong privacy guarantees \cite{tao2021benchmarking}. These marginal-based synthesizers often follow the common template of `select-measure-generate' to generate synthetic data. The `select' step involves choosing which marginals to measure from the data. The `measure' step adds noise to perturb the measured marginals to preserve privacy. Finally, the `generate' step employs algorithms to estimate a data distribution from the perturbed marginal measurements, and to sample synthetic data that closely preserves these measured marginals.


\vspace{0.5em}
\noindent
\textbf{Problem.} Existing SDG approaches operate in \textit{the centralized paradigm}, i.e.~they assume that the real data needed to train the synthesizer resides with one data holder, or, if the data originates from different data holders, that the latter are able to send their data to a trusted aggregator who in turn will use it as input for SDG algorithms. Large amounts of valuable data however are under the control of data holders (companies, hospitals, biomedical research institutes etc.) who cannot show their data to each other or to a central aggregator without raising privacy concerns. This is the bottleneck that we address, namely \textit{how to generate synthetic data based on the combined data from distributed data holders who are not able to disclose their data}. 
This includes data that is horizontally distributed, such as data across different hospitals who each have their own patients, as well as data that is vertically distributed such as in scenarios where a patient's data is distributed across multiple entities (hospitals, companies, labs, etc.). In addition to the cross-silo scenarios described above, our proposed solution would also make a scenario practical where millions of users (patients) could provide their data to produce a synthetic dataset in a way that private, individual data would never be exposed in the clear to any entity -- practically implementing a `synthetic data as a service' model. 


There is a substantial gap in the literature on solutions for SDG from distributed data sources.
While approaches such as \cite{xin2020private,behera2022fedsyn}, which are based on federated learning (FL) \cite{kairouz2021advances}, seem to provide a viable solution for this problem, with the caveat that they work either for horizontal (HFL) or for vertical (VFL) data partitioning, not both. Moreover, they 
rely on a single honest-but-curious aggregator, hence a single point of failure. Decentralized paradigms such as peer-to-peer solutions that remove the need of central entity, adapt approaches based on local differential privacy (LDP) \cite{kasiviswanathan2011can} or distributed differential privacy \cite{kairouz2021distributed} which are known to result in utility loss \cite{pasquini2023security}. 
\citealp{ramesh2023decentralised} proposed to use secure multi-party computation combined with specialized hardware (a trusted execution environment) to generate synthetic data from distributed sources.
To the best of our knowledge, there are currently no solutions that can generate differentially private synthetic data from distributed data sources using marginal-based synthesizers that work for (a) arbitrary data partitioning (horizontal, vertical, or mixed), (b) provide utility at par with that of the centralized paradigm (c) do not rely on a single trusted entity and (d) do not need specialized hardware. 

\vspace{0.5em}
\noindent
\textbf{Our Contributions.}
We observe that SDG techniques based on the `select-measure-generate' template only need to perform computations on the private data in the `select' and `measure' steps. The outputs of these
steps are protected with DP guarantees. Computations thereafter, such as the `generate' step, are \textit{post-processing}\footnote{Outputs of differentially private mechanisms are immune to post-processing \cite{dwork2014algorithmic}. } steps.
Following this observation, it is sensible to (1) secret share the data, (2) carry out the `select' and `measure' steps with differential privacy in secure multi-party computation protocols (DP-in-MPC) over the secret shared data, and (3) publish the perturbed marginals for consumption by the `generate' step. 

We capture this idea in a framework called $\gfm$ (Fig. \ref{fig:intro}) to generate differentially private synthetic tabular data from distributed data holders that works for any kind of data partitioning (horizontal, vertical, or mixed).  $\gfm$ uses MPC to provide \textit{input privacy} 
by replacing the trusted entity from the centralized paradigm with MPC servers. Unlike in the traditional central paradigm, these computing servers that execute the MPC protocols never see the private data of the data holders in the clear. To provide \textit{output privacy} guarantees, $\gfm$ uses differential privacy as in the centralized paradigm. The crucial difference however is that perturbations to provide DP are performed \textit{within} the MPC protocols,
effectively having the MPC servers simulate the role of a trusted curator implementing differential privacy mechanisms. This `DP-in-MPC' approach achieves utility at par with the centralized paradigm as shown in our experiments. 

We designed $\gfm$ to be modular. The individual algorithms for each of the select, measure, and generate steps can be replaced. Thus, the MPC protocols for the select and measure steps can be replaced or switched. This means that $\gfm$ can be easily adapted to any SDG algorithm by simply replacing or designing the MPC protocols for the given select and measure steps. Moreover, the underlying MPC primitives in the protocols can also be replaced with primitives for different adversarial models and schemes or with much more efficient primitives in the future. To illustrate this, we apply $\gfm$ to create privacy-preserving collaborative counterparts of the state-of-the-art SDG algorithms MWEM+PGM \cite{pmlr-v97-mckenna19a} and AIM \cite{mckenna2022aim}. We design MPC protocols for the select and measure steps of these algorithms and develop pipelines to generate tabular synthetic data. We evaluate these pipelines for performance, measuring runtime and communication cost. We assess the quality of the synthetic data by reporting the average workload error on benchmark datasets.
We also assess the utility of the generated synthetic data for downstream  machine learning (ML) tasks by reporting AUC and F1 scores obtained with logistic regression and random forest models trained on the benchmark data. Our evaluations 
demonstrate that $\gfm$ generates differentially private synthetic data of the same level of quality as in the centralized paradigm, while in addition providing input privacy, making it suitable for collaborative SDG from distributed data holders.  
We summarize our main contributions as:
\begin{itemize}[leftmargin=*,noitemsep,topsep=0pt]
    \item We are the first to adapt a `DP-in-MPC' approach to generate synthetic data from arbitrarily partitioned private data sources. 
    \item We propose a framework, $\gfm$, to generate differentially private synthetic tabular data from distributed private data sources. Our framework is applicable to any data distribution without the need for a trusted entity or specialized hardware. This framework achieves utility as in the centralized paradigm.
    \item We design MPC protocols for the select and measure steps of two state-of-the art marginal based synthetic data generators (MWEM+PGM and AIM), and showcase the applicability of $\gfm$ to create privacy-preserving versions of these SDG algorithms.
    \item We demonstrate the effectiveness of $\gfm$ via utility and performance experiments over benchmark datasets.
\end{itemize}

This paper surpasses a prior non-archival workshop paper in which we 
explored an early DP-in-MPC version of an older SDG algorithm 
\cite{hardt2010multiplicative} on horizontally distributed data.

\section{Preliminaries}
\label{sec:prelims}

Consider a dataset $D$ with $n$ instances. $D$ has a set of $d$ attributes denoted by $x = \{x_1,x_2,\ldots,x_d\}$.
The domain for the attribute $x_i$ is finite and is given by $\Omega_i$, i.e.~$|\Omega_i| = \omega_i$. The domain for $x$ is the cartesian product of the domains of the individual attributes, i.e.~$\Omega = \prod_{i=1}^{d} \Omega_i$. 

\vspace{0.2em}
\noindent
\textbf{Synthetic Data Generation.}
By a marginal on a set of attributes $q \subseteq x$ we mean a histogram computed over $D$ for the attributes in $q$, i.e.~it  
counts the number of occurrences 
in $D$ for each $t \in \Omega_q$ where $\Omega_q = \prod_{x_i \in q}\Omega_i$. A $k$-way marginal is a marginal computed on $k$ attributes, i.e.~$|q|=k$. 
We denote the result of the computation of a marginal on $q$ over $D$ as $\mu_q(D)$, which is basically a vector of counts.\footnote{We consider $\mu_q(D)$ to be a flattened vector for any k-way marginal.}


In what follows, we refer to a set of attributes as a \textit{query}. Marginal-based SDG algorithms aim to maintain the values of marginals for a predefined set of queries $Q$ (called a \textit{workload}) when computed over the synthetic data versus over the original data.
The algorithm typically repeats three steps in an iterative fashion: select, measure, and generate synthetic data $\Hat{D}$. In the 
the select step the algorithm usually takes the answers to the workload $Q$ on real data and intermediate synthetic data, i.e.~$\{\mu_q(D) \,|\, q \in Q\}$ and $\{\mu_q(\hat{D}) \,|\, q \in Q\}$ respectively, and selects the query $q_s$ that gives the maximum error in a differentially private manner.  
The measure step computes the noisy answer $\Hat{\mu}_{s} = \mu_s(D) + N(.)$ where $N(.)$ is the noise added to satisfy DP. The generate step 
takes the noisy answer $\Hat{\mu}_{s}$ as input 
to provide an new estimate of $\Hat{D}$.





\vspace{0.2em}
\noindent
\textbf{Secure Multiparty Computation (MPC).}
MPC is an umbrella term for cryptographic approaches that enable private computations among mutually distrustful parties \cite{cramer2000general,evans2018pragmatic}. In this paper, we follow the ``MPC as a service'' paradigm, where data holders delegate their computation to a set of non-colluding, but otherwise non-trusted servers. Moreover, we will work with secret shared-based MPC protocols. In such protocols, each data holder maps their secret inputs $z$ in secret shares according to the given MPC scheme in use. We denote secret shares of $z$ as $[\![z]\!]$. The data holders then distribute the shares to the set of servers which acts as the computing parties in MPC. Though $z$ can be reconstructed when all shares are combined, none of the MPC servers by themselves learn anything about the value of $z$. 
These servers run MPC protocols to jointly perform computations over the secret shares, in our case computations for generating synthetic data. As all computations are done over the secret shares, so that the servers do not learn the values of the inputs nor of intermediate results, i.e.~MPC provides \textit{input privacy}. 


\vspace{0.1em}
\noindent
\textit{Threat Model.} MPC protocols are designed to prevent and detect attacks by an adversary corrupting one or more parties to learn private information or to cause the result of the computation to be incorrect. The adversary -- which we assume to be static -- can have different levels of adversarial power. In the \textit{semi-honest} model, even corrupted parties follow the instructions of the protocol, but the adversary attempts to learn private information from the internal state of the corrupted parties and the messages that they receive. 
MPC protocols that are secure against semi-honest or \textit{``passive''} adversaries prevent such leakage of information. In the \textit{malicious} adversarial model, the corrupted parties can arbitrarily deviate from the protocol specification. Providing security in the presence of malicious or \textit{``active''} adversaries, i.e.~ensuring that no such adversarial attack can succeed, comes at a higher computational cost than in the passive case. 

\textit{Chosen MPC Schemes.} The protocols that we propose in Sec.~\ref{sec:method} are sufficiently generic to be used in settings with passive or active adversaries. This is achieved by changing the underlying MPC scheme to align with the desired security setting. We evaluate our protocols in an honest-majority 3-party computing setting out of which at most one party can be corrupted (3PC) \cite{araki2016high,dalskov2021fantastic}, an honest-majority 4-party computing setting with one corruption (4PC) \cite{dalskov2021fantastic}, and a dishonest-majority 2-party computation setting where each party can only trust itself (2PC) \cite{cryptoeprint:2018:482}.
%

We now give a brief introduction to MPC where we closely follow Section 3 in \cite{pentyala2021privacy}.

\textit{Data Representation in MPC.} MPC works for inputs defined over a finite field or ring. Since inputs for synthetic data set generation algorithms are finite precision real numbers, we convert all of our inputs to fixed precision \cite{FC:CatSax10} and map these to integers modulo $q$, i.e.,~to the ring $\mathbb{Z}_q =\{0,1,\ldots,q-1\}$, with $q$ a power of 2. In fixed-point representations with $a$ fractional bits, each multiplication generates $a$ additional fractional bits. To securely remove these extra bits, we use the deterministic truncation protocol by \cite{dalskov2019secure}. 

\textit{Replicated sharing-based 3PC.} To give the reader an understanding of how MPC protocols work, we give a brief description of a specific MPC protocol based on replicated secret sharing with three computing parties \cite{araki2016high}. A private value $x$ in $\mathbb{Z}_q$ is secret shared among servers (parties) $S_1, S_2,$ and $S_3$ by sselecting uniformly random shares $x_1, x_2, x_3 \in \mathbb{Z}_q$ such that 
$x_1 + x_2 +x_3 =  x \mod{q}$,
and giving $(x_1,x_2)$ to $S_1$, $(x_2,x_3)$ to $S_2$, and $(x_3,x_1)$ to $S_3$. We denote a secret sharing of $x$ by $[\![x]\!]$. 

From now on, servers will compute on the secret shares of $x$ rather than on $x$ itself. In order to proceed with the computation of a function, we need a representation of such function as a circuit consisting of addition and multiplication gates. The servers will compute the function gate by gate, by using specific protocols for computing addition of a publicly known constant to a secret shared value, addition of two secret shared values, multiplication of a secret shared value times a public constant, and multiplication of two secret shared values. After all the gates representing the function have been evaluated, the servers will hold secret shares of the desired final result of the computation. 

The three servers are capable of performing operations such as the addition of a constant, summing of secret shared values, and multiplication by a publicly known constant by doing local computations on their respective shares. 

To multiply secret shared values $[\![x]\!]$ and $[\![y]\!]$, we can use the fact that $x \cdot y=(x_1 + x_2 +x_3)(y_1 + y_2 +y_3) \mod{q}$. This means that $S_1$ computes $z_1=x_1 \cdot y_1+x_1 \cdot y_2+x_2 \cdot y_1 \mod{q}$, $S_2$ computes $z_2=x_2 \cdot y_2+x_2 \cdot y_3+x_3 \cdot y_2 \mod{q}$, and $S_3$ computes $z_3=x_3 \cdot y_3+x_3 \cdot y_1+x_1 \cdot y_3 \mod{q}$. The next step is for the servers to obtain an additive secret sharing of $0$ by choosing random values $u_1,u_2,u_3$ such that $u_1 + u_2 +u_3 = 0 \mod{q}$. This can be done using pseudorandom functions. Each server $S_i$ then computes $v_i=z_i+u_i \mod{q}$. Finally, $S_1$ sends $v_1$ to $S_3$, $S_2$ sends $v_2$ to $S_1$, and $S_3$ sends $v_3$ to $S_2$. This allows the servers $S_1, S_2$, and $S_3$ to obtain the replicated secret shares $(v_1,v_2)$, $(v_2,v_3)$, and $(v_3,v_1)$, respectively, of the value $v=x \cdot y$.


This protocol can be proven secure against honest-but-curious adversaries. In such case, corrupted players follow the protocol instructions but try to obtain as much knowledge as possible about the secret inputs from the protocol messages, and locally stored information. We can adapt this protocol to be secure even in the case of malicious adversaries. Those can arbitrarily deviate from the protocol in order to break its privacy. For the malicious case, we use the MPC scheme  proposed by \cite{dalskov2021fantastic} as implemented in MP-SPDZ \cite{cryptoeprint:2020:521}.



\vspace{0.2em}
\noindent
\textbf{Differential Privacy (DP).}
Consider two neighboring datasets $D$ and $D'$
that differ in a single instance, i.e.~$D'$ can be obtained either by adding or removing an instance from $D$ or vice-versa. A randomized algorithm $F$ is differentially private if  it generates similar output probability distributions on $D$ and $D'$~\cite{dwork2014algorithmic}. This implies that the presence or absence of an instance does not affect the output probability distribution of $F$, thus providing \textit{output privacy} by limiting the amount of information that the output reveals about any
instance. Formally, a randomized algorithm $F$ is called $(\epsilon, \delta)$-DP if for all pairs of neighboring sets $D$ and $D'$, and for all subsets $O$ of $F$'s range,
\vskip -0.2in
\begin{equation}\label{DEF:DP}
\mbox{Pr}(F(D) \in O) \leq e^{\epsilon} \cdot \mbox{Pr}(F(D') \in O) + \delta.
\end{equation}
where $\epsilon$ is the privacy budget or privacy loss and $\delta$ is the probability of violation of privacy. 
The smaller these values, the stronger the privacy guarantees. 



%
%
\section{CaPS: Collaborative and Private SDG}
\label{sec:method}


\noindent
\textbf{Problem Formulation.}
We consider a set of $N$ honest data holders $\mathbf{H} = \{{H_1},{H_2},\ldots,{H_N}\}$. Each data holder ${H_i}$ holds a dataset $D_i$ such that\footnote{We do not make any assumptions on how exactly the data is partitioned among the data holders, i.e.~each data holder may have (partial) rows and/or (partial) columns of $D$.} $D = \bigcup_{i=1}^{N} D_i$. A marginal-based synthesizer $F$ takes a workload $Q$ and $\Omega$ as input and outputs DP tabular synthetic data $\Hat{D}$. $F$ is as defined for the centralized paradigm. Our goal is to generate $\Hat{D}$ in the distributed setting with $\mathbf{H}$. 
To do so, we consider a set of $M$ independent MPC servers $\mathbf{S} = \{S_1, S_2,\ldots, S_M\}$ that perform computations over distributed private data. 


Algorithm \ref{alg:framework_gen} outlines the flow of our proposed framework $\gfm$. The key idea is to identify the steps of the SDG algorithm $F$ that require computations across data from multiple data holders 
and invoking the MPC servers to perform such computations. If the output of these computations is differentially private, the outputs can be revealed and the other steps can be done in-the-clear to avoid the overhead due to MPC.

\begin{algorithm}[tb!]
   \caption{\gfm: Generating tabular synthetic data with DP-in-MPC in the select-measure-generate template} 
   \label{alg:framework_gen}
   \textbf{Input:}  Domain $\Omega$, Queries $Q$\\
   \textbf{Output:} Synthetic Data $\Hat{D}$\\
   \textbf{Parameters:} Privacy parameters $\epsilon, \delta$\\
   \textbf{Hyper-Parameters:} Hyper-parameters for $F$\\
\begin{algorithmic}[1]
   \STATE $\mathbf{H}$: $[\![\mu^i_q]\!], [\![D_i]\!] \leftarrow$ LOCAL\_COMPUTE(\,)
   \STATE Initialization by $S_1$: $\Hat{D} \leftarrow$ INIT(\,) 
   \STATE \textbf{COMPUTE ANSWERS}: $[\![\mu_q]\!] \leftarrow \pCOMP([\![\mu^i_q]\!], [\![D_i]\!])$
   \REPEAT
   \STATE \textbf{SELECT}: $q_s \leftarrow \pSEL([\![\mu_q]\!], \Hat{\mu}_q)$ \small{//$S_1$: $\Hat{\mu}_q$ on $\Hat{D} $ } \\ 
   \STATE \textbf{MEASURE}: $\Hat{\mu}_s \leftarrow \pMSR([\![\mu_{q_s}]\!], q_s)$
    \STATE \textbf{GENERATE} by $S_1$: $\Hat{D} \leftarrow F(\Hat{\mu}_s,q_s)$
    \UNTIL{terminate}
\end{algorithmic}
\end{algorithm}

\subsection{Setup Phase}
The framework begins with a setup phase 
in which the MPC servers $\mathbf{S}$ load and compile the MPC protocols for the chosen $F$, the dimensions of the dataset $D$, and the MPC scheme. 
On Line 1 in Alg.~\ref{alg:framework_gen}, the data holders $\mathbf{H}$ perform local computations as defined by {\tt LOCAL\_COMPUTE()}; see below for a brief description. We consider $S_1$ to be the initiating server that is responsible for generating $\Hat{D}$ and publishing the generated synthetic data to the required stakeholder. $S_1$ begins with randomly initializing $\Hat{D}$ as defined by {\tt INIT()} on Line 2. We note that this initialization is the same as in the centralized version of $F$.

\noindent
\textbf{Description of {\tt LOCAL\_COMPUTE()}.}
Each data holder locally computes answers to the queries in the workload $Q$ to the best of its ability. We denote the answers computed by ${H_i}$ as $\mu^i_q, \forall q\in Q$. 
If data holder $H_i$ does not have all the attributes contained in $q$, then $H_i$ is unable to compute a local answer for $q$, so
$\mu^i_q$ is assigned a vector of 0s. All the data holders then convert their local answers into secret shares as per the given secret-sharing scheme 
as $[\![\mu^i_q]\!]$. In a non-horizontal distribution scenario, each ${H_i}$ additionally secret shares their private local dataset $D_i$ as $[\![D_i]\!]$. 



The setup phase is followed by the training phase where the MPC servers $\mathbf{S}$ run the MPC protocols for $F$ to generate synthetic data $\Hat{D}$ (Lines 3 -- 8 in Alg.~\ref{alg:framework_gen}). This phase in general does not require the $N$ data holders to stay online or participate in the generation process. However, to reduce MPC overhead, one can optimize $\gfm$ depending on the availability of any of the data holders (See Section \ref{sec:method_msr}).
The training phase proceeds with invoking MPC protocols for computing query answers $\mu_q$ over the (secret shared) real data ($\pCOMP$), and then repeatedly selecting a query $q_s$ ($\pSEL$) that needs attention and measuring/preparing a noisy answer $\Hat{\mu}_{s}$ ($\pMSR$) that can be disclosed.


\subsection{Computation of Answers on Distributed Data}\label{sec:comp}
The MPC servers run $\pCOMP$ 
to compute secret shares of $[\![\mu_q]\!]$ (Line 3 in Alg.~\ref{alg:framework_gen}). In an arbitrary setting, $\pCOMP$ can compute $[\![\mu_q]\!]$ by directly performing computations on $[\![D_i]\!], \forall i$. Such computations do not require data holders to perform local computations of answers, but does increase the MPC overhead.  
To optimize MPC overhead, we propose for each query $q$ that all the data holders who have all attributes in $q$ perform one-time local computations to compute partial query answers $[\![\mu^i_q]\!]$ (see Line 1 in Alg.~\ref{alg:framework_gen}); note that for queries involving only one attribute, i.e.~one-way marginals, many such local computations can be done. The first step in  $\pCOMP$  is then to  simply add the secret shares of the locally computed answers from each data holder $[\![\mu^i_q]\!], \forall q \in Q$ (Lines 1--5 in Prot.~\ref{prot:comp}). This is then followed by computing secret shares of workload answers $[\![\mu_{q^*}]\!]$ on $[\![D_i]\!]$ for queries $q^*$ that can not be computed locally, i.e.~queries containing attributes that are distributed across different data holders. We call the set of these queries $Q^*$.
After the execution of $\pCOMP$, the MPC servers $\mathbf{S}$ hold the secret shares of $[\![\mu_q]\!], \forall q \in Q$.


Note that when the data is horizontally partitioned, $Q^* = \emptyset$, and the workload answers $[\![\mu_q]\!]$ can be fully computed by adding secret shares of locally computed results, i.e.~Lines 1--5 in Prot.~ \ref{prot:comp}.
This results in a total of $(N-1) \cdot |Q| \cdot \max(\omega_q)$ additions in MPC implying that  $\pCOMP$ is linear in the number of data holders for a given $Q$ and $\Omega$ for horizontally distributed data.
When $Q^* \neq \emptyset$, secret sharings of the complete query answers can be obtained by letting the servers $\mathbf{S}$ perform more computation over the secret-shared data of the data holders, as explained below.



\begin{myprotocol}
   \caption{$\pCOMP$:\\ MPC Protocol for \textbf{COMPUTE ANSWERS}}
   \label{prot:comp}
    \textbf{Input:} Secret shares of locally computed answers $[\![\mu^i_q]\!]$, Secret shares of data $[\![D_i]\!]$ (needed if not horizontal),
   Queries $Q$, including queries $Q^*$ with attributes that are distributed among data holders, 
   Domain $\Omega$  \\    
    \textbf{Output:} $[\![\mu_q]\!], \forall q \in Q$ 
   \begin{algorithmic}[1]
   \FOR{$i$ = $1$ to $N$}
   \FORALL{$q \in Q$}
   \STATE $[\![\mu_q]\!]$ $\leftarrow$ $[\![\mu_q]\!]$ + $[\![\mu^i_q]\!]$
   \ENDFOR 
   \ENDFOR
   \IF{$Q^* \neq \emptyset$}
       \STATE $[\![D]\!]$ $\leftarrow \pJOIN([\![D_i]\!] | i = {1 \ldots N} )$
   \ENDIF
    \FORALL{ $q^* = \{a_1,a_2\}$ in $Q^*$}
    \FOR{$i$ = $1$ to $n$}
    \STATE $[\![x]\!]$ $\leftarrow$ $[\![D_{a_1}[i]]\!]$
    \STATE $[\![y]\!]$ $\leftarrow$ $[\![D_{a_2}[i]]\!]$
    \FOR{$j$ $\in \Omega_{a_1}$ and $k$ $\in \Omega_{a_2}$}
    \STATE $[\![m]\!] \leftarrow \pMUL(\pEQ([\![x]\!],j) ,\pEQ([\![y]\!],k))$
    \STATE $[\![\mu_{q^*}[j*|\Omega_{a_2}| + k]]\!] \leftarrow [\![\mu_{q^*}]\!] + [\![m]\!]$
    \ENDFOR
    \ENDFOR
    \ENDFOR
   \STATE \textbf{return} $[\![\mu_q]\!]$, $\forall q \in Q$
\end{algorithmic}
\end{myprotocol}

\noindent
\textbf{MPC protocol $\pCOMP$.}
Prot.~\ref{prot:comp} computes $[\![\mu_q]\!]$ and is optimized for AIM and MWEM+PGM in an arbitrary distributed setting. The workload $Q$ in these algorithms consists of 
1-way and 2-way marginals.
Lines 1--5 aggregate the workload answers from all the data holders by performing addition of secret shares of $[\![\mu^i_q]\!]$. 
This aggregation accounts for computation of 1-way marginals for any arbitrary setting and 2-way marginals for the horizontal setting.


If $Q^* \neq \emptyset$, then $\mathbf{S}$ proceed to assemble a secret-sharing of the joint dataset using the protocol $\pJOIN$ on Line 7. $\pJOIN$ initializes a matrix $[\![D]\!]$ of dimensions $n \times d$. We assume that all the data samples are aligned before $\pJOIN$.\footnote{This can be achieved using protocols for Private Set Intersection (PSI) available in literature.} $\pJOIN$ then combines all the secret-sharings $[\![D_i]\!]$ into a secret-sharing $[\![D]\!]$ of the overall dataset $D$ using simple assignment statements. In Protocol 2, $\pJOIN$ requires $n.d$ assignment operations and works for any number of data holders.
$\mathbf{S}$ then run Lines 9--18 to compute secret-sharings of 2-way marginals.
Consider a 2-way marginal $q^* \in Q^*$ which is represented as a pair $q^* = \{a_1,a_2\}$. $[\![D_{a_1}]\!]$ denotes the secret shares of $[\![D]\!]$ for the $a_1$ attribute and $[\![D_{a_2}]\!]$ for $a_2$. 
On Lines 10--17, the MPC servers iterate over all the $n$ data samples in $[\![D_{a_1}]\!]$ and $[\![D_{a_2}]\!]$ to compute the number of occurrences of all the combinations of values in the domains $\Omega_{a_1}$ and $\Omega_{a_2}$. Line 14 in Prot.~\ref{prot:comp} relies on MPC primitives for multiplication $\pMUL$ and equality testing $\pEQ$ to check if a combination of attribute values occurs in the data. For details about these primitives, see Appendix \ref{app:mpc}. On Line 14, the MPC servers compute $[\![m]\!]$ which is a secret sharing of 0 or 1 that adds to the the number of occurrences of the combination of attributes values held in $[\![\mu_{q^*}]\!]$.
We note that a major MPC overhead is due to Line 14 performed in the loop. 
This indicates that for an arbitrary data setting,  there are $(N-1) \cdot |Q| \cdot \max(\omega_q) + n \cdot \max(\omega_{q})^2 \cdot |Q^*|$ additions, $2 \cdot n \cdot \max(\omega_{q})^2 \cdot |Q^*|$ equality checks, and $n \cdot \max(\omega_{q})^2 \cdot |Q^*|$ multiplications. Given a fixed set of queries $Q$ and fixed domain $\Omega$, Protocol \ref{prot:comp} grows linearly in the total number of samples $n$ and the total number of data holders $N$.

MPC protocols are typically designed as specific circuits, composed of a sequence of addition and multiplication gates. Therefore, changing the functionality to be implemented in MPC usually requires the design of a new circuit. Protocol 2 is specifically designed to suit algorithms  that consider 1-way and 2-way marginals. One can extend Protocol 2 to compute $p$-way marginals as discussed in Appendix \ref{sec:prot2ext}.
We consider optimizing the protocol for different scenarios as future work.

\begin{myprotocol}
\small{
   \caption{$\pSEL$: MPC Protocol for \textbf{SELECT}}
   \label{prot:sel}
    \textbf{Input:} Secret shares of $[\![\mu_q]\!]$, Estimated answer $\Hat{\mu_q}$,  length of domains $\omega_q$ for each $q$, privacy parameter $b$, sensitivity $w$ \\    
    \textbf{Output:} Selected query $q_s$
   \begin{algorithmic}[1]
   \STATE $[\![err]\!]$ $\leftarrow \pERR([\![\mu_q]\!], \Hat{\mu_q})$  
   \STATE $[\![err]\!]$ $\leftarrow \pNORM([\![err]\!])$ 
   \STATE $[\![prob]\!]$ $\leftarrow \pPROB([\![err]\!], b, w, \omega_q)$
   \STATE $[\![s]\!]$ $\leftarrow \pRC([\![prob]\!])$ 
   \STATE Reveal $s$ 
   \STATE \textbf{return} $q_s$
\end{algorithmic}
}
\end{myprotocol}

\subsection{Selection of the Query}
In the select step, $\mathbf{S}$ run the MPC protocol $\pSEL$ to select the query $q_s$ based on which the estimate of $\Hat{D}$ should be improved (Line 5 in Alg.~\ref{alg:framework_gen}). We note that $\pSEL$ is independent of the data distribution setting as the set of MPC servers $\mathbf{S}$ already have secret shares of the computed answers $[\![\mu_q]\!]$ from all the data holders. 
Prot.~\ref{prot:sel} provides a general template for selecting $q_s$. $S_1$ who holds $\Hat{D}$ computes $\Hat{\mu}_q$ and provides it as input to Prot.~\ref{prot:sel}.
On Line 1 the MPC servers compute a secret-sharing of the error $[\![err]\!]$ between $[\![\mu_q]\!]$ and $\Hat{\mu}_q$ by taking the sum of absolute differences of the answers using $\pERR$.  This is followed by computing secret shares of the normalized errors $[\![err]\!]$ 
using $\pNORM$ on Line 2. A secret shared probability vector over the queries is then constructed using $\pPROB$ on Line 3 which takes the secret-shared normalized errors $[\![err]\!]$, the privacy parameters -- the scale $b$ and sensitivity $w$ -- 
as input. We describe the MPC subprotocols $\pERR$, $\pNORM$ and $\pPROB$ in Appendix \ref{app:mpc}. What happens within these protocols may differ from one SDG algorithm to the next; 
we discuss these subprotocols specifically for AIM and MWEM+PGM in protocols \ref{prot:sel_aim} and \ref{prot:sel_mwem} respectively in Appendix A. 
Finally on Line 4, $\mathbf{S}$ run subprotocol $\pRC$ to randomly select the query using exponential mechanism, thus implementing the DP-in-MPC paradigm. $\pRC$ outputs the secret shares of the randomly chosen index $[\![s]\!]$, which is revealed to $S_1$. The query with index $s$ is selected as $q_s$. The MPC overhead due to $\pSEL$ is in the order of $|Q| \times \max(\omega_q)$ of the computations involved. This means that $\pSEL$ does not depend on the number of data holders $N$ or the total number of samples $n$ but on he given number of queries $|Q|$ and the domain of the dataset $\Omega$.

\noindent
\textbf{MPC protocol for $\pRC$.}
Protocol \ref{prot:sel_rc} takes as input the secret shares of the computed probabilities, $[\![prob]\!]$ and chooses the first index of the probability vector for which the probability value satisfies a condition based on random threshold. Lines 1--8 compute the random threshold $[\![$t$]\!]$ using the MPC primitives for multiplication ($\pMUL$), random number generation ($\pRDM(0,1)$). Lines 10--15 select the index conditioned on the threshold without exiting the loop, as it could reveal the value of returned index to the adversary. We design Lines 10--15 to prevent such side-channel attacks.
To understand this part of the code, note that we have a list $p[1..|Q|]$ of non-decreasing values, i.e.~the cumulative probability sums, and the MPC servers have to find the first index $i$ in $p[1..|Q|]$ for which $p[i] >$ t. In a mock example with $|Q|=10$, and assuming that the first such $p[i]$ value is at position 7, the tests on Line 11 will generate the results 0,0,0,0,0,0,1,1,1,1. On Line 12, these results are accumulated in $k$, which eventually becomes 4, and the desired index is computed as $N-(k-1)=10-3=7$. Lines 14--15 take care of the edge case when $p[i] \leq $ t for all $i$ (i.e.~$k$ is $0$). We protect the value of $k$ by employing MPC primitives for multiplication to simulate a conditional statement.



\begin{myprotocol}
   \caption{$\pRC$: MPC Protocol for random selection}
   \label{prot:sel_rc}
    \textbf{Input:} Secret shares of probability vector $[\![prob]\!]$, length of probability vector $|Q|$\\    
    \textbf{Output:} Secret shares of selected index 
    $[\![s]\!]$
   \begin{algorithmic}[1]
   \STATE sum $\leftarrow$ $0$
   \STATE Initialize a vector $\bold{p}$ of length $N$\\
   \FOR{$i$ = $1$ to $|Q|$}
   \STATE $[\![$sum$]\!]$ $\leftarrow$ $[\![$sum$]\!]$ + $[\![prob[i]]\!]$
   \STATE $[\![p[i]]\!]$ $\leftarrow$ $[\![$sum$]\!]$
   \ENDFOR
   \STATE $[\![$r$]\!] \leftarrow \pRDM(0,1)$ // with protocol for random number generation $\pRDM$\\
   \STATE $[\![$t$]\!]$ $\leftarrow$ $\pMUL([\![$sum$]\!],[\![$r$]\!])$
   \STATE $k$ $\leftarrow$ $0$ \\
   \FOR{$i$ = $1$ to $|Q|$}
   \STATE $[\![$c$]\!]$ $\leftarrow$  $\pGT([\![p[i]]\!]$, $[\![$t$]\!])$
   \STATE $[\![k]\!]$ $\leftarrow$ $[\![k]\!]$ + $[\![$c$]\!]$
   \ENDFOR
   \STATE $[\![$c$]\!]$ $\leftarrow$ $\pEQ([\![k]\!], 0)$
   \STATE $[\![s]\!]$ $\leftarrow$ $N - \pMUL([\![k]\!] - 1, 1-[\![$c$]\!])$
   \STATE \textbf{return} $[\![s]\!]$
\end{algorithmic}
\end{myprotocol}






\subsection{Measuring Answer to the Selected Query }\label{sec:method_msr}
In the measure step, $\mathbf{S}$ run the MPC protocol $\pMSR$ to compute the noisy answer to selected query $q_s$ (Line 6 in Algorithm \ref{alg:framework_gen}). As $\mathbf{S}$ holds the secrets shares $[\![{\mu}_{s}]\!]$  computed for any arbitrary data distribution,  $\pMSR$  computes $[\![\Hat{\mu}_{s}]\!]$
by generating the secret shares of noise vector  $[\![{\gamma}]\!]$ and adding it to the $[\![{\mu}_{s}]\!]$. Once the $\mathbf{S}$ compute $[\![\Hat{\mu}_{s}]\!]$, it is revealed to $S_1$ for generation step.

The MPC overhead is mainly due to generation of $[\![{\gamma}]\!]$. To reduce MPC overhead in an arbitrary distributed setting, one can consider the availability of the data holder who holds the attributes defined in $q_s$  and request for generation of the noisy vector ($S_1$ plays the role of sending request and receiving response). This is a direct optimization in a vertical distributed setting when the selected query is a 1-way marginal. However, the advantage of generating noise in MPC is that the noise is added in a correct and private manner, such that private inputs cannot be reconstructed back. One can also optimize the MPC primitives to generate the secret shares of noise as we consider in Protocol \ref{prot:msr}.

Protocol \ref{prot:msr} computes Gaussian noise as required by AIM and MWEM+PGM for arbitrary distributed setting. To do so, we consider Irwin-Hall approximation to generate Gaussian samples on Lines 2 -- 7 as it reduces the MPC overhead considerably. The protocol for generating noise requires ($13.\omega_{s} + \max(\omega_q)$) additions and $12.\omega_{s}$ random bit generations. These lines can be replaced by other MPC protocols for other sampling techniques such as Box-Muller method as we show in the Prot. \ref{prot:msr_bm} in Appendix \ref{app:mpc}.


\begin{myprotocol}
\small{
   \caption{$\pMSR$: MPC Protocol for \textbf{MEASURE} using Gaussian noise }
   \label{prot:msr}
    \textbf{Input:} Secrets shares of $[\![\mu_{q_s}]\!]$, length of domain $\omega_{s}$ for $q_s$, scale $b$ \\    
    \textbf{Output:} Noisy measurement $\Hat{\mu}_{s}$
   \begin{algorithmic}[1]
   \STATE Initialize vector $[\![\gamma]\!]$ of length $\max(\omega_r)$ with $0$s
   \FOR{i = 0 to $\omega_{q_s}$}
   \STATE $[\![sum]\!] \leftarrow 0$
   \FOR{j = 0 to 12}
   \STATE $[\![sum]\!] \leftarrow [\![sum]\!] + \pRDM(0,1)$
   \ENDFOR
   $[\![\gamma[i]]\!] \leftarrow [\![sum]\!] - 6$
   \ENDFOR
   \STATE $[\![\Hat{\mu}_{s}]\!]  \leftarrow [\![\mu_{s}]\!] + b.[\![\gamma]\!]$
   \STATE \textbf{return} $\Hat{\mu}_{s}$
\end{algorithmic}
}
\end{myprotocol}

\subsection{Generation of Synthetic Data}
$S_1$ takes the noisy measurement $\Hat{\mu}_{s}$ and runs the estimate algorithm of $F$ to generate $\Hat{D}$. $S_1$ can run this step without the need of MPC protocols as it takes DP input which means that the privacy of $D$ is preserved due to the post-processing property of DP.




\begin{table}[h!]
    \caption{\textbf{Utility evaluation}. Synthetic data was generated with $\epsilon=1.0$. For the distributed scenario with $\gfm$, $N = 2$ and $M = 3$ (3PC passive \cite{araki2016high}). 
    All values are averaged across 3 runs. Abbreviations: CDP = Central Differential Privacy with trusted aggregator (no input privacy); $\gfm$ = Our proposed approach for arbitrary distribution; H = Horizontal distribution; V = Vertical distribution; LR = Logistic Regression; RF = Random Forest.} 
    \vskip 0.15in
    \centering
    \small{
    \begin{sc}
        \begin{tabular}{ c l | ccc }
        \toprule
        & &    \multicolumn{3}{c}{Breast-cancer (Categorical)} \\
        & &  {CDP} & {$\gfm$}{(H)} & {$\gfm$}{(V)}\\
        \midrule   
        \parbox[t]{0.6mm}{\multirow{5}{*}{\rotatebox[origin=c]{90}{\textbf{AIM}}}} &
                 LR-AUC  & 0.49 & 0.49 &  0.51   \\
               &  RF-AUC  & 0.58 &0.45  & 0.53    \\ 
               &  LR-F1  & 0.47 & 0.43  & 0.48    \\ 
             &  RF-F1  & 0.54 & 0.44 & 0.51   \\ 
            &  error $\Delta$  & 0.30 & 0.23 &  0.23  \\

        \midrule
        \parbox[t]{0.6mm}{\multirow{5}{*}{\rotatebox[origin=c]{90}{\textbf{\scriptsize{MWEM+PGM}}}}}  
                & LR-AUC  & 0.50 & 0.55 &  0.44  \\
               & RF-AUC  & 0.51 &  0.54 & 0.49   \\ 
                & LR-F1  & 0.48  & 0.49 & 0.43   \\ 
              & RF-F1  & 0.44 & 0.46 &  0.50  \\ 
        
              & error $\Delta$  &0.24  & 0.21 &  0.21 \\

        \midrule
        & & \multicolumn{3}{c}{COMPAS (Categorical)} \\
         & &  {CDP} & {$\gfm$}{(H)} & {$\gfm$}{(V)}\\
        \midrule   
        \parbox[t]{0.6mm}{\multirow{5}{*}{\rotatebox[origin=c]{90}{\textbf{AIM}}}} 
                & LR-AUC  & 0.67 & 0.66 & 0.68  \\
               &  RF-AUC  &0.65  & 0.65 & 0.67  \\ 
               &  LR-F1  & 0.62 & 0.62  &  0.63    \\ 
             &  RF-F1    & 0.61 & 0.61 & 0.62  \\ 
              & error $\Delta$  & 0.017 & 0.019 & 0.015   \\

        \midrule
        \parbox[t]{0.6mm}{\multirow{5}{*}{\rotatebox[origin=c]{90}{\textbf{\scriptsize{MWEM+PGM}}}}}  
                & LR-AUC & 0.66 & 0.66 &  0.66  \\
               &  RF-AUC & 0.62 & 0.61  & 0.64   \\ 
               &  LR-F1  & 0.62 & 0.60 &  0.61  \\ 
             &  RF-F1   & 0.59 & 0.58 & 0.60   \\ 
 
              & error $\Delta$  & 0.022 & 0.022 &  0.022  \\

        \midrule
        & &  \multicolumn{3}{c}{Diabetes (Continuous)} \\
           & &  {CDP} & {$\gfm$}{(H)} & {$\gfm$}{(V)}\\
        \midrule   
        \parbox[t]{0.6mm}{\multirow{5}{*}{\rotatebox[origin=c]{90}{\textbf{AIM}}}} 
                & LR-AUC  & 0.76 &0.77  & 0.73  \\
               &  RF-AUC & 0.74 & 0.72 & 0.71  \\ 
               &  LR-F1 & 0.68 & 0.65 & 0.66   \\ 
             &  RF-F1   & 0.66 & 0.63 & 0.63  \\ 
              & error $\Delta$  & 0.14 & 0.13  & 0.13  \\

        \midrule
        \parbox[t]{0.6mm}{\multirow{5}{*}{\rotatebox[origin=c]{90}{\textbf{\scriptsize{MWEM+PGM}}}}}  
                & LR-AUC & 0.62  & 0.64 & 0.67  \\
               &  RF-AUC  & 0.57 & 0.60 & 0.63  \\ 
               &  LR-F1 & 0.55 & 0.52 & 0.60  \\ 
             & RF-F1   & 0.55 & 0.56 & 0.58  \\  
              & error $\Delta$  & 0.15 & 0.14 &  0.14 \\

        \bottomrule
        \end{tabular} 

        \end{sc}
    }
    \label{tab:results_utility1}
\end{table}

\subsection{Note on Modularity}
The protocols $\pSEL$ and $\pMSR$ can be replaced by 
custom MPC protocols for 
the select and measure steps in one's SDG algorithm of choice. In our work, we specifically design MPC protocols for AIM and MWEM+PGM. Our choice of these algorithms is based on the literature showcasing that these are the state-of-the-art synthetic tabular data generation techniques \cite{mckenna2022aim,pereira2023assessment}. 

$\gfm$ is applicable to workload-based synthetic generation algorithms too such as RAP \cite{vietri2022private} that follow the select-measure-generate template. Our $\pMSR$ protocol can be used as is for other algorithms such as RAP and MST \cite{mckenna2021winning}. If the considered distribution setting is only horizontal, $\pMSR$ can benefit in terms of MPC overhead by employing distributed DP in case of Gaussian noise addition. We also notice that $\pCOMP$ is equivalent to computing joint histograms from all the data holders. This protocol can be replaced in our framework with other efficient protocols, if available, for the given scheme and datasets distributions (e.g.~\cite{bell2022distributed,asharov2023secure}).

The implementations of MPC protocols for differentially private mechanisms can also be replaced to satisfy different project requirements. For example, the Gaussian sampling protocol $\pGSS$, which is based on the Box-Muller transform, can be replaced by other methods \cite{canonne2020discrete}.


%
%
\section{Experimental Evaluation}
\label{sec:results}
\noindent
\textbf{Datasets.}
We evaluate $\gfm$ on three datasets: breast-cancer \cite{misc_breast_cancer_14}, prison recidivism (COMPAS)\footnote{\url{https://www.propublica.org/datastore/dataset/compas-recidivism-risk-score-data-and-analysis}} \cite{compas}, and diabetes \cite{smith1988using}. We randomly split all the datasets into train and test in an 80\% to 20\% ratio.
The breast-cancer dataset has 10 categorical attributes and 285 samples. 
The COMPAS data consists of categorical data. We utilize the same version as 
in \cite{calmon2017optimized}, which consists of 7 categorical features and 7,214 samples.
The diabetes dataset has 9 continuous attributes and 768 samples. 
We use the train sets to generate synthetic data and the test sets to evaluate the quality of the synthetic data.

\noindent
\textbf{Metrics.} 
To assess utility, we train classifiers
on the generated synthetic data and test the models on the real test data. For breast-cancer, the task is to predict if the cancer will recur. For COMPAS, the task is to predict whether a criminal defendant will re-offend. For the diabetes dataset, the task is to classify a patient as diabetic. We train logistic regression and random forest models on the generated synthetic datasets and report the AUC-ROC and F1 score. 
We also evaluate  $\gfm$ for statistical utility of the generated data as  
the workload error $\Delta$  \cite{mckenna2022aim}:
\vskip -0.2in
\begin{equation*}
    \Delta(D,\Hat{D}) = \frac{1}{|Q|} \sum_{q \in Q} \| \mu_{q}(D) -  \mu_{q}(\Hat{D}) \| 
\end{equation*}

\noindent
\textbf{Evaluation Setup.}
We implemented the MPC protocols of $\gfm$ in MP-SPDZ \cite{keller2020mp}.
We evaluate $\gfm$ for horizontal ($\gfm$(H)) and vertical ($\gfm$(V)) data distribution scenarios, and compare against the centralized paradigm (CDP) in which all data holders give their data to a trusted aggregator (i.e.~no input privacy). 
For the horizontal setup, we distribute the samples randomly and evenly among the data holders.
For the vertical setup, we distribute the attributes randomly and evenly among the data holders.

\noindent
\textbf{Utility Analysis.}
Table \ref{tab:results_utility1} shows that $\gfm$ can generate synthetic data whose utility is at par with the CDP in terms of ML utility and statistical utility for both horizontal and vertical partitions. For the COMPAS and diabetes datasets, CDP and $\gfm$ give similar ML and statistical utility. We attribute the higher variability observed in the cancer data experiments (AUC and F1) to the small data size. The variation observed in results in Table \ref{tab:results_utility1} are due to to randomness resulting from DP.

\begin{table}[!ht]
    \caption{\textbf{Runtime evaluation}. Synthetic data was generated with $\epsilon=1.0$. For the distributed scenario with $\gfm$, $N = 2$ and $M = 3$ (3PC passive \cite{araki2016high}). For $\gfm$ we report runtimes for experiments done in a \textbf{\textit{simulated environment}}. All values are in seconds and averaged across 3 runs. Abbreviations: CDP = Central Differential Privacy with trusted aggregator (no input privacy); $\gfm$ = Our proposed approach for arbitrary distribution; H = Horizontal distribution; V = Vertical distribution.
    } 
    \vskip 0.15in
    \centering
    \begin{sc}
    \scalebox{0.80}{
    \small{
        \begin{tabular}{ c ll  | cccc }
        \toprule
         & &  & $|Q|\times \max(\omega_q)$ & {CDP} &  $\gfm$(H) & $\gfm$(V) \\
              \midrule
        & &  &    \multicolumn{4}{c}{Breast-Cancer ($n=228,d=10$)} \\
        \midrule

                \multicolumn{2}{c}{AIM}& & 57$\times$77 & 98.065 & 99.789 & 98.992\\
        \multicolumn{2}{c}{MWEM+PGM}& &45$\times$77 &81.187 & 82.905  &  100.295 \\

        \midrule
                & &  &  \multicolumn{4}{c}{COMPAS ($n=4120,d=9$)}  \\
      \midrule
                \multicolumn{2}{c}{AIM} &  & 45$\times$9 & 153.383 & 155.559 & 301.390\\
        \multicolumn{2}{c}{MWEM+PGM} &  & 36$\times$9 &61.477 & 61.656 & 152.153  \\

            \midrule
                & &  & \multicolumn{4}{c}{Diabetes ($n=614,d=9$)} \\
        \midrule
  
                \multicolumn{2}{c}{AIM} & & 45$\times$25 & 97.222 & 113.909 & 97.403 \\
        \multicolumn{2}{c}{MWEM+PGM}& & 36$\times$25 &60.926 & 63.727 & 91.760 \\

        \bottomrule
        \end{tabular} 
        }
        }
        \end{sc}
    \label{tab:results_runtimes1}
\end{table}

\noindent
\textbf{Performance Analysis.}
To measure the average time to generate synthetic data in CDP and $\gfm$, we run experiments in a simulated environment on Azure D8ads\_v5 8 vCPUs, 32Gib RAM. We leverage MP-SPDZ to create the simulated environment in a single machine. In Table \ref{tab:results_runtimes1}, we observe that, as expected, $\gfm$ takes longer than CDP due to MPC overhead. For MWEM+PGM, it takes longer to generate synthetic data in a vertical setup due to overhead of $\pCOMP$. For AIM, the vertical setup takes comparatively less time. This is because AIM requires computation of one-way marginals before the selection step, which can be done by the data holders in a vertical distributed scenario. 
In the horizontal setup, computation of one-way marginals has to be done in MPC, generating additional overhead in computation of marginals compared to the vertical setup. 
We also note that $|Q|\times \max(\omega_q)$ impacts the runtime for ($\gfm$(H)), whereas $|Q|\times \max(\omega_q)$ and $n$ impact the runtime for ($\gfm$(V)). We believe these runtimes are acceptable since fast response time is not crucial for SDG. 
Moreover, the benefits of generating synthetic data while preserving both input and output privacy surpass this additional cost.

In Table \ref{tab:results_mpc_eval}, we evaluate individual MPC subprotocols for different threat models when they are run on independent instances of Azure Standard F16s v2 (16 vcpus, 32 GiB memory) and network bandwidth of 12.5Gbps. The results are in-line with the literature \cite{keller2022secure}. We observe that 3PC passive provides the least MPC overhead. For $t$ iterations of the loop in Alg.~\ref{alg:framework_gen}, the additional time due to MPC in a 3PC setting is (2.882 + $t \cdot $1.1913) sec for AIM and (2.882 + $t \cdot$ 0.0463) sec for MWEM+PGM for the chosen parameters.\footnote{We note that the number of parties in MPC corresponds to $M$ servers in our framework and not $N$ data holders.} Given that the one-time additional cost is comparatively much lower than the time  it takes to generate synthetic data itself in-the-clear, $\gfm$ achieves near-practical performance. With further optimizations of MPC primitives in the future, $\gfm$ has the capability to be deployed in real-world scenarios. Our results demonstrate a clear path for future research in this direction and adapting various synthetic data generation techniques smartly in the ``DP-in-MPC'' paradigm.   
The major MPC overhead due to the arbitrary distributed setting in $\gfm$ is attributed to the computations in $\pCOMP$. The performance of this protocol is impacted by the total number of samples $n$ in $D$. We empirically evaluate the scalability for $n$ and $N$ in Appendix \ref{app:perf}.

\begin{table}[h!]
    \caption{\textbf{Performance evaluation of MPC protocols for different threat models}. MPC protocols are run with $N=2, n=614, d=9, |Q|=45, \max(\omega_q)=25$. We run experiments with $M=2, 3, 4$ in a LAN setup and the mentioned threat models. We report runtimes in seconds and total communication cost (Comm.) in MB for the online phase of the MPC protocols. $\pSEL$(A) refers to the MPC protocol for `select' for the AIM algorithm and $\pSEL$(M) for the MWEM+PGM algorithm. $\pCOMP$ refers to the computation of 1- and 2-way marginals in a vertical distribution.} 
    \vskip 0.15in
    \centering
    \begin{sc}
    \scalebox{0.89}{
    \small{

        \begin{tabular}{ l | rr rr }
        \toprule
        Protocol & \multicolumn{2}{c}{2PC passive} & \multicolumn{2}{c}{3PC passive}  \\
               & time(s) & Comm.(MB) & time(s) & Comm.(MB)  \\
        \midrule
        $\pMSR$    & 0.020 & 2.214 & 0.0043 & 0.261   \\
        $\pSEL$(A) & 2.972 & 119.053 & 1.187 & 0.554  \\
        $\pSEL$(M) & 0.783 & 167.658 & 0.042 &  5.341 \\
        $\pCOMP$   & 135.482 & 30137.40 & 2.882 & 696.644 \\        
       \midrule
        Protocol  &  \multicolumn{2}{c}{3PC active} & \multicolumn{2}{c}{4PC active} \\
               & time(s) & Comm.(MB) & time(s) & Comm.(MB) \\
        \midrule
        $\pMSR$    & 0.030 & 2.140 & 0.0157 &  0.714 \\
        $\pSEL$(A) & 1.773 & 11.971 & 2.169 & 13.707 \\
        $\pSEL$(M) & 0.306 & 33.946 & 0.135 & 4.700  \\
        $\pCOMP$   & 21.612 & 3830.53 & 6.00 &  1615.61 \\        
       \bottomrule
  \end{tabular} 
        }
        }
        \end{sc}
    \label{tab:results_mpc_eval}
\end{table}




%
%

%
%
\section{Related Work}
\label{sec:realted}




To the best of our knowledge, there are only a handful of works on generating synthetic data from multiple private data sources \cite{pflgan,behera2022fedsyn,zhao2023libertas,ramesh2023decentralised,ghavamipour2023federated,tajeddine2020privacy}. The work closest to ours is \cite{maddock2024flaim} that focuses on loosely-coordinated federated settings rather than synchronized horizontal distributed setting. All of these methods either work only for a given data distribution scenario or rely on specialized hardware. 
In contrast with existing work, we focus on generating synthetic data with the ``DP-in-MPC'' paradigm for SDG algorithms that follow the select-measure-generate template, in a manner that works for any arbitrary distribution and does not require specialized hardware. In the general modular framework that we have introduced to this end, one can 
plug in other protocols from the MPC literature, including protocols for sampling noise, see e.g. \cite{wei2023securely,pettai2015combining,zhao2019facct}.


%
%
\section{Conclusion}
\label{sec:concl}
We introduced a general framework $\gfm$ that enables collaborative and private generation of tabular synthetic data based on real data from multiple data holders. 
$\gfm$ follows the select-measure-generate template to  generate synthetic data regardless of how the real data is distibuted among the data holders, i.e.~horizontally, vertically, or mixed.
We leverage the idea of ``DP-in-MPC'' to provide input privacy by performing computations on secret shared data, and applying differential privacy mechanisms within MPC itself. Letting MPC servers emulate a trusted central entity allows us to provide the same level of output privacy and utility as in the centralized paradigm, however without having to sacrifice input privacy.
We demonstrated the applicability of $\gfm$ for the state-of-the-art marginal based synthetic data generators AIM and MWEM+PGM. 
We consider generating synthetic data using ``DP-in-MPC'' for other data modalities and their state-of-the art generation algorithms as a future research direction.   

\newpage
\section*{Acknowledgements}
The authors would like to thank Marcel Keller for making the MP-SPDZ framework available. The authors thank Microsoft for the generous donation of cloud computing credits through the UW Azure Cloud Computing Credits for Research program. Sikha Pentyala is a Carwein-Andrews Distinguished Fellow and is supported by a JP Morgan Chase PhD fellowship, the UW Global Innovation Fund and the UWT Founders Endowment. This research was, in part, funded by the National Institutes of Health (NIH) Agreement No. 1OT2OD032581. The views and conclusions contained in this document are those of the authors and should not be interpreted as representing the official policies, either expressed or implied, of the NIH.

\section*{Impact statement}
Our work advocates for responsible use of sensitive and private data while benefiting different tasks of data science. In our paper, we propose a general framework that generates synthetic data from multiple sources. We believe that synthetic data is a multi-advantageous solution for data sharing while mitigating privacy concerns. Our work contributes to generation of synthetic data based on training data from multiple private data sources, thereby improving the quality and diversity of generated data. We envision a scenario where our framework can be used to generate synthetic data in sensitive domains such as healthcare. 
Organizations can use the generated synthetic data to augment their local datasets, perform data imputations or train effective and fair ML models.
We build upon the existing state-of-the art algorithms to generate synthetic data. Our proposed framework is limited by the disadvantages and affects of these existing algorithms (such as fairness and robustness of generated synthetic data, and vulnerability to adversarial privacy attacks). Users must be aware of such limitations/disadvantages and validate the suitability of the algorithm to generate synthetic data when considering our framework. We additionally note that our framework provides privacy and utility at the cost of runtime and communication, so choosing the framework for time-sensitive scenarios is left to the user's discretion.
In our experiments we use publicly available datasets thereby adhering to the ethical guidelines for data usage.

\bibliography{ref}
\bibliographystyle{icml2024}

\newpage
\appendix
\onecolumn

\section{MPC protocols.}\label{app:mpc}
\noindent
\textbf{MPC primitives.}
The MPC schemes listed in Section \ref{sec:prelims} provide a mechanism for the servers to perform cryptographic primitives through the use of secret shares, namely addition of a constant, multiplication by a constant, and addition of secret shared values, and multiplication of secret shared values (denoted as $\pMUL$). Building on these cryptographic primitives, MPC protocols for other operations have been developed in the literature. 
We use \cite{cryptoeprint:2020:521}:
\begin{itemize}[leftmargin=*,noitemsep,topsep=0pt]

    \item Secure random number generation from uniform distribution $\pRDM$ : In $\pRDM$, each party generates $l$ random bits, where $l$ is the fractional precision of the power 2 ring representation of real numbers, and then the parties define the bitwise XOR of these $l$ bits as the binary representation of the random number jointly generated.
    
    \item Secure random bit generation $\pRDMB$ : In $\pRDMB$, each party generates the secret share of a single random bit, such that the generated bit is either 0 or 1 with a probability of 0.5.

    \item Secure multiplication $\pMUL$ : At the start of this protocol, the parties have secret sharings $[\![a]\!]$ and $[\![b]\!]$; at the end, then they have a secret share of $c=a .b$. 
 
    \item Secure equality test $\pEQ$ : At the start of this protocol, the parties have secret sharings $[\![a]\!]$; at the end if $a = 0$, then they have a secret share of $1$, else a secret sharing of $0$.

    \item Secure less than test $\pLT$ : At the start of this protocol, the parties have secret sharings $[\![a]\!]$ and $[\![b]\!]$ of integers $a$ and $a$; at the end of the protocol they have a secret sharing of $1$ if $a < b$, and a secret sharing of $0$ otherwise.
    
    \item Secure greater than test $\pGT$ : At the start of this protocol, the parties have secret sharings $[\![a]\!]$ and $[\![b]\!]$ of integers $a$ and $b$; at the end of the protocol they have a secret sharing of $1$ if $a > b$, and a secret sharing of $0$ otherwise.

    \item Secure greater than test $\pGTE$ : At the start of this protocol, the parties have secret sharings $[\![a]\!]$ and $[\![b]\!]$ of integers $a$ and $b$; at the end of the protocol they have a secret sharing of $1$ if $a \geq b$, and a secret sharing of $0$ otherwise.

    \item Other primitives : We use secure maximum protocol ($\pMAX$), secure exponential protocol ($\pEXP$) and secure logarithm protocol ($\pLN$) as the building blocks for our protocols. $\pLN$ uses the polynomial expansion for computing logarithm and $\pEXP$ in turn uses the $\pLN$ to compute exponential. $\pMAX$ inherently uses the $\pGT$ repeatedly over a list by employing variant of Divide-n-Conquer approach. At the start of all of these primitives, parties hold the secret sharings $[\![x]\!]$ and at the end of the protocol they hold the secret shares of the corresponding computed values.
    
    \item Other generic subprotocols : $\pERR$ computes the secret shares of error between secrets shares of two vectors. $\pNORM$ normalizes, either by scaling or computing L1 norm, the secret shares of the error vector. $\pPROB$ computes the probabilities for a given set of secret shares of vectors. $\pSQRT$ computes secret shares of the square root of $[\![x]\!]$, $\pSIN$ computes secret shares of the sine of $[\![x]\!]$, $\pCOS$ computes secret shares of the cosine of $[\![x]\!]$. $\pSUM$ computes the sum of secret shares of a given vector. $\pSOFTMAX$ computes the softmax for a given vector. (See \cite{keller2020mp})
\end{itemize}
MPC protocols can be mathematically proven to guarantee privacy and correctness. We follow the universal composition theorem that allows modular design where the protocols remain secure even if composed with other or the same MPC protocols \cite{canetti2000security}.



\noindent
\textbf{Description of $\pSEL$ for AIM.}
Protocol \ref{prot:sel_aim} is the straight forward implementation of the select step from the centralized algorithm of AIM. Lines 2--7 compute the secrets shares of errors between the answers from real and synthetic data  and form the subprotocol $\pERR$. This protocol relies on $\pNORML1$ to compute the L1-norm of the errors. Protocol \ref{prot:aim_l1norm} computes secret shares of L1-norm of the error vector. This protocol takes the input size m which is the length of the answer for for query $q_i$, i.e.~m $= \omega_{q_i}$ (this is an implementation details where we consider the unpadded vector answer).
Lines 9--12 normalize the secret shares of errors by scaling it with secret share of maximum error computed using primitive $\pMAX$. Lines 9--12 form the subprotocol for normalize $\pNORM$.
Line 14 computes secret shares of the probabilities by relying on the subprotocol $\pSOFTMAX$. This is defined as 
$\pPROB$ for AIM. Finally on Line 16, MPC protocol for exponential mechanism is called $\pRC$ that outputs the selected query based on computed secret shares of the probabilities.

\begin{myprotocol}[h!]
\small{
   \caption{$\pSEL$: MPC Protocol for \textbf{SELECT} for AIM}
   \label{prot:sel_aim}
    \textbf{Input:} Secrets shares of $[\![\mu_q]\!]$, Estimated answer $\Hat{\mu}_q$,  length of domains $\omega_q$ for each $q$, privacy parameters $\epsilon$, a vector bias $b$ for each $q$, weight for queries $w$ , max\_sensitivity $s$ \\    
    \textbf{Output:} Selected query $q_s$
   \begin{algorithmic}[1]
   \STATE ********* Compute errors - $\pERR$ *********
   \STATE Initialize a vector ${err}$ of length $max(\omega_q)$\\
   \FOR{$i=1$ to $|Q|$}
   \STATE $[\![$diff$]\!]$ $\leftarrow$  $[\![\mu_{q_i}]\!] - \Hat{\mu}_{q_i}$\\
   \STATE $[\![err[i]]\!]$ $\leftarrow$ $\pNORML1([\![$diff$]\!], \omega_{q_i})$// computes L1 norm of the errors
   \STATE $[\![err[i]]\!]$ $\leftarrow$ $\pMUL(w[i], [\![err[i]]\!] -b[i])$ 
   \ENDFOR
   \STATE ********* Normalize errors - $\pNORM$ *********
   \STATE $[\![$max\_err$]\!]$ $\leftarrow$ $\pMAX([\![{err}]\!])$
   \FOR{$i=1$ to $|Q|$}
   \STATE $[\![err[i]]\!]$ $\leftarrow$ $[\![err[i]]\!] - [\![$max\_err$]\!]$
   \ENDFOR
   \STATE ********* Compute probabilities - $\pPROB$ *********
   \STATE $[\![prob]\!]$ $\leftarrow$ $\pSOFTMAX($0.5  $\cdot \epsilon \cdot (1/s) \cdot [\![err\!]$)
   \STATE ********* Select random index - $\pRC$ *********
   \STATE $[\![s]\!]$ $\leftarrow \pRC([\![prob]\!])$ 
   \STATE Reveal $s$
   \STATE \textbf{return} $q_s$
\end{algorithmic}
}
\end{myprotocol}

\begin{myprotocol}[h!]
\small{
   \caption{$\pNORML$: MPC Protocol to compute L1-norm}
   \label{prot:aim_l1norm}
    \textbf{Input:} Input vector $[\![$diff$]\!]$, length of vector m \\    
    \textbf{Output:} Normalized vector $[\![err[i]]\!]$
   \begin{algorithmic}[1]
   \STATE sum $\leftarrow$ 0
   \FOR{$i=1$ to m}
    \STATE $[\![$sign$]\!]$ $\leftarrow$  $\pLT([\![$diff[i]$]\!], 0)$ 
    \STATE $[\![$abs\_diff$]\!]$ $\leftarrow$ $\pMUL(1 - 2 \cdot [\![$sign$]\!], [\![$diff[i]$]\!])$ 
   \STATE $[\![$sum$]\!]$ $\leftarrow$  $[\![$sum$]\!]$ + $[\![$abs\_diff$]\!]$\\
   \ENDFOR
   \STATE \textbf{return} $[\![$sum$]\!]$
\end{algorithmic}
}
\end{myprotocol}

\begin{myprotocol}[h!]
\small{
   \caption{$\pSEL$: MPC Protocol for \textbf{SELECT} for MWEM+PGM}
   \label{prot:sel_mwem}
    \textbf{Input:} Secrets shares of $[\![\mu_q]\!]$, Estimated answer $\Hat{\mu}_q$,  length of domains $\omega_q$ for each $q$, privacy parameters $\epsilon$ \\    
    \textbf{Output:} Selected query $q_s$
   \begin{algorithmic}[1]
   \STATE ********* Compute errors - $\pERR$ *********
   \STATE Initialize a vector ${err}$ of length $max(\omega_q)$\\
   \FOR{$i=1$ to $|Q|$}
   \STATE $[\![$diff$]\!]$ $\leftarrow$  $[\![\mu_{q_i}]\!] - \Hat{\mu}_{q_i}$\\
   \STATE $[\![$sign$]\!]$ $\leftarrow$  $\pLT([\![$diff$]\!], 0)$ 
   \STATE $[\![$abs\_diff$]\!]$ $\leftarrow$ $\pMUL(1 - 2 \cdot [\![$sign$]\!], [\![$diff$]\!])$ // computes absolute difference of error between answers on real data and synthetic data
   \STATE $[\![$sum$]\!]$ $\leftarrow$ $\pSUM(1 - 2 \cdot [\![$abs\_diff$]\!])$
   \STATE $[\![err[i]]\!]$ $\leftarrow$ $[\![$sum$]\!] - \omega_{q_i}$
   \ENDFOR
   \STATE ********* Normalize errors - $\pNORM$ *********
   \STATE $[\![$max\_err$]\!]$ $\leftarrow$ $\pMAX([\![{err}]\!])$
   \FOR{$i=1$ to $|Q|$}
   \STATE $[\![err[i]]\!]$ $\leftarrow$ $[\![err[i]]\!] - [\![$max\_err$]\!]$
   \ENDFOR
   \STATE ********* Compute probabilities - $\pPROB$ *********
   \STATE $[\![prob]\!]$ $\leftarrow$ $\pSOFTMAX($0.5  $\cdot \epsilon \cdot [\![err\!]$)
   \STATE ********* Select random index - $\pRC$ *********
   \STATE $[\![s]\!]$ $\leftarrow \pRC([\![prob]\!])$ 
   \STATE Reveal $s$
   \STATE \textbf{return} $q_s$
\end{algorithmic}
}
\end{myprotocol}

\noindent
\textbf{Description of $\pSEL$ for MWEM+PGM.}
Protocol \ref{prot:sel_mwem} is the straight forward implementation of the select step from the centralized algorithm of MWEM+PGM. Lines 2--9 compute the secrets shares of errors between the answers from real and synthetic data and form the subprotocol $\pERR$. Lines 11--14 normalize the secret shares of errors by scaling it with secret share of maximum error computed using primitive $\pMAX$. Lines 11--14 form the subprotocol for normalize $\pNORM$.
Line 16 computes secret shares of the probabilities by relying on the subprotocol $\pSOFTMAX$. This is defined as 
$\pPROB$ for MWEM+PGM. Finally on Line 18, MPC protocol for exponential mechanism is called $\pRC$ that outputs the selected query based on computed secret shares of the probabilities.

\noindent
\textbf{Description of $\pMSR$ using Box-Muller Method.} Lines 3--6 in Protocol \ref{prot:msr} can be replaced by the subprotocol \ref{prot:msr_bm} which relies on the transform by \cite{box1958note} to generate samples of the Gaussian unitary distribution, namely $\lceil \omega_s/2 \rceil$ pairs of Gaussian samples. 
For each pair, the MPC servers securely generate secret shares of two random number $u$ and $v$ uniformly distributed in [0,1] using MPC primitive $\pRDM$
on Lines 3--4. On Lines 5--8, the MPC servers then compute a secret sharing of $\sqrt{-2 \ln(u)} \cdot \cos (2 \pi v)$ and of $\sqrt{-2 \ln(u)} \cdot \sin (2 \pi v)$ using MPC protocols for $\pSQRT$, $\pSIN$, $\pCOS$, and $\pLOG$ (\cite{cryptoeprint:2020:521}).In case $d$ is odd, one more sample needs to be generated. The parties do so on Lines 11--12 by executing $\pGSS$ to sample a vector of length 2 and only retain the first coordinate.

\begin{myprotocol}[h!]
   \caption{$\pGSS$: Box-Muller to generate Guassian sample with mean 0 and variance 1}
   \label{prot:msr_bm}
    \textbf{Input:} Vector length $\omega_s$.\\    
    \textbf{Output:} A secret-shared vector $[\![\gamma]\!]$ of length $\omega_s$ sampled from Gaussian distribution with mean 0 and variance 1
   \begin{algorithmic}[1]
   \STATE Initialize vector $[\![\gamma]\!]$  of length $d$ 
    \FOR{$i=0$ {\bfseries to} $\omega_s/2$} 
        \STATE $[\![u]\!] \leftarrow \pRDM(0,1)$ 
        \STATE $[\![v]\!] \leftarrow \pRDM(0,1)$ 
        \STATE $[\![r]\!] \leftarrow \pSQRT(-2 \pLN([\![u]\!]))$ 
        \STATE $[\![\theta]\!] \leftarrow 2 \pi [\![v]\!]$ 
        \STATE $[\![s_{2i}]\!] \leftarrow \pMUL([\![r]\!], \pCOS([\![\theta]\!]))$ 
        \STATE $[\![x_{2i+1}]\!] \leftarrow \pMUL([\![r]\!], \pSIN([\![\theta]\!]))$ 
        
    \ENDFOR
    
    \IF{$d$ is odd}
    \STATE $[\![p]\!] \leftarrow \pGSS(2)$ 
    \STATE $[\![s_{d-1}]\!] \leftarrow [\![p_{0}]\!]$ 
    \ENDIF
    \STATE \textbf{return} {$[\![\gamma]\!]$}
\end{algorithmic}
\end{myprotocol}

\noindent
\textbf{Description of $\pSEL$ with Laplacian noise generation.} $\pMSR$ can be replaced by Protocol \ref{prot:msr_lap} when one needs to add Laplacian noise.  The noise is sampled as  $b \cdot \ln ${\,}l$ \cdot $ c where $b$ is the scale, l is a random value drawn from the uniform distribution in [0,1] and c is a random value selected from $\{-1,1\}$. 
On Lines 3--4, the MPC servers straightforwardly compute l and its natural log. 
To compute c, the parties, on line 5, generate secret shares of a random bit $[\![$r$]\!]$, i.e.~ a value $\in  \{0,1\}$ is chosen, where each value has a chance of 50\% to be chosen. On line 6, the parties transform r to a value $\in \{-1,1\}$ using the logic c $=2 \cdot$ r $- 1$. Lines 7--10 is straightforward computation of the noise vector $[\![\gamma[i]]\!]$ and noisy measurement $[\![\Hat{\mu}_{s}]\!]$.

One can also replace the lines 3--6 by the following MPC pseudocode. This pseudocode drawns Laplace noise based on $l = - b \cdot \sgn(u) \cdot \ln(1-2|u|)$, where $u$ is a random variable drawn from a uniform distribution in $[-0.5,0.5]$. $l$ has distribution $Lap(0,b)$. This follows from the inverse cumulative distribution function for $Lap(0,b)$.

$[\![u]\!] \leftarrow \pRDM(-0.5,0.5)$\\ 
$[\![\mbox{sgn}_u]\!] \leftarrow \pGTE([\![u]\!],0)$\\ 
$[\![\mbox{abs}_u]\!] \leftarrow \pMUL([\![u]\!],[\![\mbox{sgn}_u]\!]))$ \\      
$[\![\mbox{ln}_u]\!] \leftarrow \pLN(1 - 2 \cdot [\![\mbox{abs}_u]\!])$\\
$[\![l]\!] \leftarrow -1 \cdot \pMUL([\![\mbox{ln}_u]\!], [\![\mbox{sgn}_u]\!])$\\
$[\![\gamma[i]]\!] \leftarrow [\![l]\!]$

\begin{myprotocol}[h!]
\small{
   \caption{$\pSEL$: MPC Protocol for \textbf{MEASURE} using Laplacian noise }
   \label{prot:msr_lap}
    \textbf{Input:} Secrets shares of $[\![\mu_{q_s}]\!]$, length of domain $\omega_{s}$ for $q_s$, scale $b$ \\    
    \textbf{Output:} Noisy measurement $\Hat{\mu}_{s}$
   \begin{algorithmic}[1]
   \STATE Initialize vector $[\![\gamma]\!]$ of length $\max(\omega_r)$ with $0$s
   \FOR{i = 0 to $\omega_{q_s}$}
   \STATE $[\![l]\!]$ $\leftarrow$  $\pRDM(0,1)$ // with protocol for random number generation $\pRDM$
   \STATE $[\![$ln\_l$]\!]$ $\leftarrow$ $\pLN([\![$l$]\!])$ // with secure logarithm protocol $\pLN$
\STATE $[\![$r$]\!]$ $\leftarrow$  $\pRDMB()$ // with protocol for random bit generation $\pRDM$\\
\STATE $[\![$c$]\!]$ $\leftarrow$  $2 \cdot [\![$r$]\!] - 1$ 
\STATE   $[\![\gamma[i]]\!] \leftarrow \pMUL([\![$ln\_l$]\!], [\![$c$]\!])$ // with secure multiplication protocol $\pMUL$\\ 
   \ENDFOR
   \STATE $[\![\Hat{\mu}_{s}]\!]  \leftarrow [\![\mu_{s}]\!] + b.[\![\gamma]\!]$
   \STATE \textbf{return} $\Hat{\mu}_{s}$
\end{algorithmic}
}
\end{myprotocol}

\paragraph{Extending Protocol 2 to compute $p$-way marginals.}\label{sec:prot2ext}
MPC protocols are typically designed as specific circuits, composed of a sequence of addition and multiplication gates. Therefore, changing the functionality to be implemented in MPC usually requires the design of a new circuit. Protocol 2 is specifically designed to suit algorithms  that consider 1-way and 2-way marginals. Our framework enables the design of new MPC protocols or the utilization of existing ones for complex scenarios, facilitating the generation of tabular synthetic data for arbitrary distributed data. 

One can extend the Protocol 2 to compute 3-way marginals in the following ways:
\begin{itemize}
    \item A direct extension of Protocol 2 to compute $p$-way marginals 
    is possible. Say the $p$ marginals are represented by a set $\mathcal{M}$. This requires $p$ conditionals instead of 2 on Line 13 of Protocol 2. Instead of 2 equality checks on Line 14, this requires $p$ equality checks. This means a total of $p \cdot n \cdot \prod_{i \in \mathcal{M}} \omega_i$ equality checks and $p \cdot \prod_{i \in \mathcal{M}} \omega_i$ multiplications. As stated on Line 328, $\pCOMP$ can be replaced by other protocols such as ~\cite{bell2022distributed}.
    
    For $m=3$, change $q^* = \{a_1,a_2,a_3\}$  and add  $[\![z]\!]$ $\leftarrow$ $[\![D_{a_3}[i]]\!]$ after Line 12 on Protocol 2. Change Line 13 to include $l$ $\in \Omega_{a_3}$ and add a multiplication operation with $\pEQ([\![z]\!],l$ on Line 14. Line 15 should be modified to index accordingly.  Please see Protocol \ref{prot:pway}.

    We will add a discussion on extending protocol 2 to compute of $m$-way marginals in naive manner. We will state that optimizing the protocol for different scenarios is considered as future work.
    
    \item Another possible way to compute 3-way marginals is by repeated computations of 2-way marginals. We compute for $\{a_1,a_2\}$ first and then get a column with the concatenated values for the feature $a_1 || a_2$ and then compute 2-way marginal for $\{a_1||a_2, a_3\}$. Depending the domain size of each column, this  might perform better or worse than the above method.
\end{itemize}

\begin{myprotocol}[ht!]
   \caption{MPC Protocol to compute $p$-way marginals}
   \label{prot:pway}
    \textbf{Input:} Secret shares of data $[\![D_i]\!]$ (needed if not horizontal),
   Queries $Q$, including queries $Q^*$ with attributes that are distributed among data holders, 
   Domain $\Omega$  \\    
    \textbf{Output:} $[\![\mu_q]\!], \forall q \in Q$ 
   \begin{algorithmic}[1]
   \IF{$Q^* \neq \emptyset$}
       \STATE $[\![D]\!]$ $\leftarrow \pJOIN([\![D_i]\!] | i = {1 \ldots N} )$
   \ENDIF
    \FORALL{ $q^* = \{a_1,a_2,\ldots,a_p\}$ in $Q^*$}
    \FOR{$i$ = $1$ to $n$}
    \STATE $[\![x]\!]$ $\leftarrow$ $[\![D_{a_1}[i]]\!]$; $[\![y]\!]$ $\leftarrow$ $[\![D_{a_2}[i]]\!]$; $\ldots$ $[\![z]\!]$ $\leftarrow$ $[\![D_{a_p}[i]]\!]$
    \FOR{$j$ $\in \Omega_{a_1}$ and $k$ $\in \Omega_{a_2}$ $\cdots$ and $l$ $\in \Omega_{a_p}$}
    \STATE $[\![m]\!] \leftarrow \pMUL(\pEQ([\![x]\!],j) ,\pEQ([\![y]\!],k), \cdots, \pEQ([\![z]\!],l))$
    \STATE index $\leftarrow$ compute the index for marginal vector
    \STATE $[\![\mu_{q^*}[$index$]\!] \leftarrow [\![\mu_{q^*}]\!] + [\![m]\!]$
    \ENDFOR
    \ENDFOR
    \ENDFOR
   \STATE \textbf{return} $[\![\mu_q]\!]$, $\forall q \in Q$
\end{algorithmic}
\end{myprotocol}

Design and optimization of every possible MPC protocols for all scenarios of select-measure-generate paradigm is beyond the scope of our work. We note that our work provides an important baseline framework that can be adapted to particular algorithms.  

\paragraph{Discussion on common datasets.}
Our framework, in principle, works for all the distributed settings due to the modularity offered. This includes scenarios where the data holders hold common data as shown in Table \ref{tab:Alicecommon}-\ref{tab:Alicecommon2}.  To illustrate the how MPC can be leveraged, Protocol 2 for $\pCOMP$ considers disjoint dataset. $\pCOMP$ also works with very little modifications when the data is common across some of the data holders as shown in Table \ref{tab:Alicecommon}-\ref{tab:Alicecommon2} for two data holders. In such case, the computations in $\pCOMP$ begin with executing $\pJOIN$ that results in a union of all datasets. This is followed by computation of all the marginals in MPC (such as computations done on Lines 9--18 of Protocol 2 to compute 2-way marginals). This requires removing Lines 1--6 in Protocol 2 and having $Q^* = Q$. This will of course result in change in the number of computations performed in MPC. 

\begin{table}[!htb]
    \begin{minipage}{.5\linewidth}
        \caption{Data held by $H_1$}
        \label{tab:Alicecommon}
      \centering
        \begin{tabular}{r|c|c}
        ID & a & b \\
        \hline
        Alice & 0 & 1 \\
         Bob & 1 & 1 \\
    \end{tabular}
    \end{minipage}%
    \begin{minipage}{.5\linewidth}
      \centering
    \caption{Data held by $H_2$}
    \label{tab:Alicecommon2}
        \begin{tabular}{r|c|c|c}
        ID & a & b & c \\
        \hline
        Alice & 0 & 1 & 2\\
        Charlie & 1 & 0 & 2 \\
    \end{tabular}
    \end{minipage} 
    
\end{table}

\paragraph{Implementing DP in MPC.}
Keeping in mind the dangers of implementing DP with floating point arithmetic \cite{mironov2012significance}, we stick with the best practice of using fixed-point and integer arithmetic as recommended by, for example, OpenDP \footnote{https://opendp.org/}. We implement all our DP mechanism using their discrete representations and use 32 bit precision to ensure correctness.

We also remark that finite precision issues can also impact the exponential mechanism. If the utility of one of the classes in the exponential mechanism collapses to zero after the mapping into finite precision, pure DP becomes impossible to achieve. We can deal with such situation by using approximate DP for an appropriate value of $\delta$. Such situation did not happen with the data sets used in our experiments.

\section{Additional experiments.}\label{app:experiments}
\subsection{Performance analysis}\label{app:perf}
We run additional experiments to evaluate the scalability of $\gfm$.

\begin{figure}
    \centering
    \includegraphics[width=\textwidth]{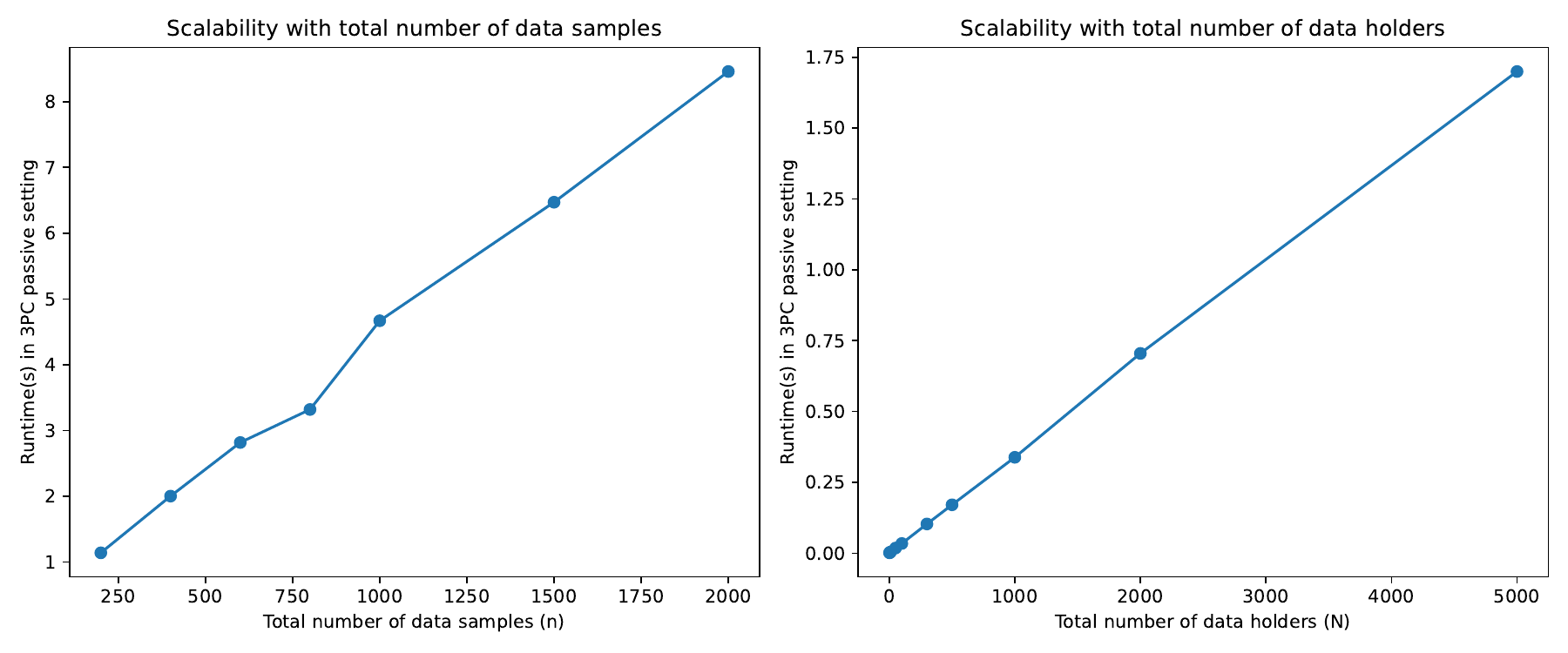}
    \caption{\textbf{Scalability of $\pCOMP$ in a 3PC passive setting}. MPC protocols are run with $M=3,d=10,|Q|=36,max(\omega_q)=25$. On left: Scalability of $\pCOMP$ for different number of total dataset size $n$. 
    On right: Scalability of $\pCOMP$ for different number of data holders $N$. }
    \label{fig:results_scale_N}
\end{figure}

\paragraph{Scalability with total training samples $n$}
We run experiments in a 3PC passive setting and evaluate the scalability of $\pCOMP$ in Figure \ref{fig:results_scale_N}. This is the only protocol that depends on the value of total number of samples over all the distributed datasets. We note that we run experiments for $\pCOMP$ which computes 1-way and 2-way marginals in a distributed setting. The overhead due to the number of data holders (requires only additions) in this case is negligible when compared to computation of 2-way marginal for large $n$. This means that our findings are independent of the number of data holders $N$.

\paragraph{Scalability with data holders $N$}
To illustrate the affect of increase in number of data holders, we consider a horizontal distributed scenario where each data holder holds equal number of samples. The overhead in this case is due to only addition of workload answers from $N$ data holders for a given $|Q|$ and $\omega$.

We note that the scalability of other protocols $\pSEL$ and $\pMSR$ depend on $|Q|$ and $\omega$. We think the impact due to $|Q|$ and $\omega$ is similar to the impact on centralized algorithms. Based on literature, runtimes for MPC protocols grow with large $M$, depending on the MPC scheme available for $M$. $\gfm$ adapts MPC-as-a-service scenario, where we can choose the optimal number of MPC server, $M$. We think based on the current literature in MPC $M=3$ is a better option for $\gfm$.




\section{Code.}
We have made the code available at \url{https://github.com/sikhapentyala/MPC_SDG/tree/icml}


\end{document}